\documentclass[preprint2]{aastex}
\usepackage{bm}
\usepackage{textcomp}
\usepackage{natbib}
\begin{document}
\title{Modeling the quantum interference signatures of the Ba~{\sc ii} D$_2$ 4554~\AA\ 
line in the second solar spectrum}

\author{H. N. Smitha$^{1}$, K. N. Nagendra$^{1}$,
J. O. Stenflo$^{2,3}$ and M. Sampoorna$^{1}$} 
\affil{$^1$Indian Institute of Astrophysics, Koramangala,
Bangalore, India}
\affil{$^2$Institute of Astronomy, ETH Zurich,
CH-8093 \  Zurich, Switzerland }
\affil{$^3$Istituto Ricerche Solari Locarno, Via Patocchi,
6605 Locarno-Monti, Switzerland}

\email{smithahn@iiap.res.in; knn@iiap.res.in; stenflo@astro.phys.ethz.ch;
sampoorna@iiap.res.in}

\begin{abstract}
Quantum interference effects play a vital role 
in shaping the linear polarization profiles of solar spectral lines.
The Ba~{\sc ii} D$_2$ line at 4554~\AA\ is a prominent example, where the 
$F$-state interference effects due to the odd isotopes produce
polarization profiles, which are very different from those of the even
isotopes that have no $F$-state interference. It is therefore
necessary to account for the contributions from the different isotopes
to understand the observed linear polarization profiles of 
this line. Here we do radiative transfer modeling with
partial frequency redistribution (PRD) of such 
observations while accounting for the interference effects and isotope composition.
The Ba~{\sc ii} D$_2$ polarization profile is found to be strongly governed by
the PRD mechanism. We show how a full PRD treatment succeeds in
reproducing the observations, while complete 
frequency redistribution (CRD) alone fails
to produce polarization profiles that have any resemblance with the
observed ones. However, we also find that the line center polarization is sensitive to 
the temperature structure of the model atmosphere. To obtain a good fit
to the line center peak of the observed Stokes $Q/I$ profile, a 
small modification of the FALX model atmosphere is needed, by lowering
the temperature in the line-forming layers. Because of the pronounced
temperature sensitivity of the Ba~{\sc ii} D$_2$ line it may not be
a suitable tool for Hanle magnetic-field diagnostics of the solar
chromosphere, because there is currently no straightforward way to separate the
temperature and magnetic-field effects from each other. 
\end{abstract}

\maketitle

\section{Introduction}
The linearly polarized spectrum of the Sun known as the 
Second Solar Spectrum exhibits signatures of a number of physical
processes not seen in the intensity spectrum and which may also be used
to diagnose weak and turbulent magnetic fields that are inaccessible
with the ordinary Zeeman effect. 
Many of the most prominent spectral lines in the Second Solar
Spectrum, like Na\,{\sc i} D$_1$ and D$_2$, \
Ba\,{\sc ii} D$_1$ and D$_2$, and Ca\,{\sc ii} H and K, are governed by
effects of quantum interference between states of different total angular
momentum ($J$ or $F$ states).
The profound importance of quantum interference for the formation of these
lines was first demonstrated both observationally and theoretically by 
\citet[][see also \citealt{1997A&A...324..344S}]{1980A&A....84...68S}.

Quantum interference between the $J$-states 
governs the linear polarization
signatures of atomic multiplets such as
the  Ca\,{\sc ii} H and K, Na\,{\sc i} D$_1$ and D$_2$, and Cr~{\sc i}
triplet lines. 
The theory of such $J$-state interference was developed to include the effects of partial
frequency redistribution (PRD) by \cite{2011ApJ...733....4S,2013JQSRT.115...46S}.
This theory was then applied successfully to model the linear polarization profiles
of the Cr~{\sc i} triplet at 5204-5208~\AA~ by \cite{2012A&A...541A..24S}.

\begin{figure*}
\centering
\includegraphics[width=12.0cm, height=6.0cm]{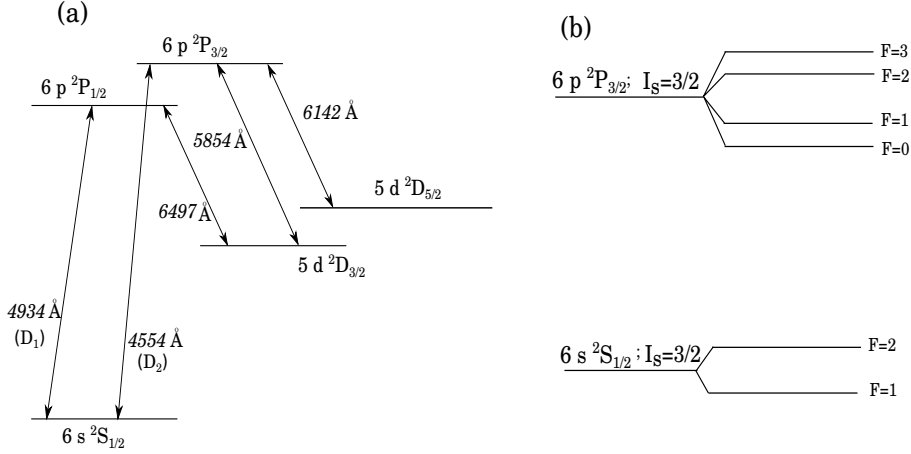}
\caption{(a) represents the Ba~{\sc ii} 
model atom for the even isotopes, while for the odd isotopes
the atomic model is modified by 
replacing two of the levels, $^2P_{3/2}$ and $^2S_{1/2}$, with their hyperfine 
structure components, as shown in (b). In (b), $I_s$ is the nuclear spin.
The energy levels are not drawn to scale.}
\label{level-diag}
\end{figure*}

When an atom possesses nuclear spin the $J$ states are split into $F$
states (hyperfine structure splitting). The quantum interference
between the $F$ states produces 
depolarization in the line core. Examples of lines governed by 
$F$-state interference are the Na\,{\sc i} D$_2$, 
Ba\,{\sc ii} D$_2$, and Sc\,{\sc ii} line at 4247\,\AA. 
The Ba~{\sc ii} D$_2$ line is due to the 
transition between the upper fine structure level $J=3/2$ and the lower level 
$J=1/2$ (see Figure~\ref{level-diag}(a)). In the odd isotopes of Ba,
both the upper and lower levels
undergo hyperfine structure splitting (HFS) due to the nuclear spin $I_s=3/2$,
resulting in four upper and two lower $F$-states
(see Figure~\ref{level-diag}(b)). The quantum interference
between the upper  
$F$-states needs to be taken into account
in the modeling of the Ba\,{\sc ii} D$_2$ line. The odd isotopes 
contribute about 18\%\ of the total Ba abundance in the Sun
~\citep[c.f. Table 3 of][]{2009ARA&A..47..481A}.
The remaining 82\%\ comes from the even isotopes, which are not subject
to HFS (because $I_s=0$). 

The intensity profile of the Ba~{\sc ii} D$_2$ line has earlier been studied 
extensively, for example by \cite{1974SoPh...39...19H} and  
\citet[][and the references cited therein]{1978SoPh...56..237R}.
Some of these studies aimed at determining the solar abundance of Ba. 
Observations with the high precision spectro-polarimeter ZIMPOL by 
\cite{1997A&A...321..927S} clearly revealed the existence of three
distinct peaks in the linear polarization $(Q/I)$ profiles of the Ba~{\sc ii} D$_2$ line. 
The nature of these peaks could subsequently be theoretically clarified 
by \cite{1997A&A...324..344S}, who used the last scattering
approximation to model the $Q/I$ profiles. It was demonstrated that the 
central $Q/I$ peak is due to the even isotopes of Ba, 
while the two side peaks are due to the odd isotopes. 

Using a similar last scattering approximation, the magnetic
sensitivity of the Ba line 
was explored by \cite{2007ApJ...666..588B}.
Both these papers however did not account for radiative transfer
or PRD effects. 
The potential of using the Ba\,{\sc ii} D$_2$ line as a 
diagnostic tool for chromospheric weak turbulent magnetic fields 
and the important role of PRD were 
discussed by \cite{2009A&A...493..201F}, but the treatment was limited
to the even Ba isotopes, for which HFS is absent. In contrast, our
radiative-transfer treatment with PRD in the present paper includes
both even and odd isotopes and the full effects of HFS with $F$-state interferences. 

The theory of $F$-state interference with PRD in the non-magnetic collisionless regime
was recently developed in \citet[][herafter P1]{2012ApJ...758..112S}. 
The PRD matrix was also incorporated into
the polarized radiative transfer equation. The transfer equation was then solved for the case of 
constant property isothermal atmospheric slabs. In the present paper we extend the work of P1 
to solve the line formation problem in realistic 1-D model atmospheres in order to model
the Ba\,{\sc ii} D$_2$ line profile observed in a quiet region close to the solar limb. 

The outline of the paper is as follows: In Section~{\ref{rt}} we present the 
polarized radiative transfer equation which is suitably modified
to handle several isotopes of Ba. In Section~{\ref{obs-details}} 
we present the details of the observations. In Section~{\ref{modeling-pro}}
we discuss the model atom and the model atmosphere used. The results are presented in 
Section~{\ref{results}} with concluding remarks in Section~{\ref{conclusions}}.

\section{Polarized line transfer equation with $F$-state interference}
\label{rt}
The polarization of the radiation field
is in general represented by the full Stokes vector $(I,Q,U,V)^{\rm T}$.
However, in the absence of a magnetic field Stokes $U$ and $V$ 
are zero in an axisymmetric 1-D atmosphere. Hence in a non-magnetic
medium the Stokes vector $(I,Q)^{\rm T}$
is sufficient to express the polarization state of the radiation field. 
The transfer equation in the reduced Stokes vector basis  \citep[see][]{2012A&A...541A..24S}
is
\begin{eqnarray}
\mu \frac{{\partial {\bm{\mathcal I}}(\lambda, \mu, z)}
}{\partial z}= 
-k_{\rm tot}(\lambda, z)
\left[{\bm{\mathcal I}}(\lambda, \mu, z) - 
{\bm{\mathcal S}}(\lambda, z)\right],
\label{transfer}
\end{eqnarray}
with positive $Q$ defined to represent linear polarization
oriented parallel to the solar limb.
The quantities appearing in Equation~(\ref{transfer}) are defined in the reference
mentioned above. However, we need to generalize the previous definitions 
of opacity and source vector to handle even and odd isotope contributions
together.

The total opacity 
$k_{\rm tot}(\lambda, z)= k_{l}(z)\phi_g(\lambda,z) + 
\sigma_c(\lambda, z) + k_{\rm th}(\lambda, z)$, where  $\sigma_c$ and 
$k_{\rm th}$ are the continuum scattering and continuum absorption
coefficients, respectively.
In the present treatment, $k_{\rm th}$ also includes the contribution from the 
blend lines, which are assumed to be depolarizing
and hence are treated in LTE.
$k_{l}$ is the wavelength 
averaged absorption coefficient for the $J_a \to J_b$ transition.
$J_a$ and $J_b$ are the electronic angular momentum quantum numbers 
of the lower and upper level, respectively.
$\phi_g$ is the Voigt profile function written as
 \begin{equation}
\phi_g(\lambda,z) = 0.822 ~\phi_e(\lambda,z) + 0.178 ~\phi_o(\lambda,z).
\label{phi-g}
\end{equation}
$\phi_e(\lambda,z)$ is the Voigt profile function for the even isotopes of Ba~{\sc ii}
corresponding to the $J_a=1/2 \to J_b=3/2$ transition 
in the absence of HFS. 
The profile function for the odd isotopes 
is $\phi_o(\lambda,z)$, which is the weighted sum of the individual
Voigt profiles $\phi(\lambda_{F_bF_a},z)$ representing 
each of the $F_a \to F_b$ absorption transitions. Here $F_a$ and $F_b$ are the total angular 
momentum quantum numbers of the initial and the intermediate hyperfine split levels,
respectively. $\phi_o(\lambda,z)$  is the same as $\phi_{\rm HFS}(\lambda,z)$
defined in Equation~(7) of P1 and is given by
\begin{eqnarray}
&&\phi_o(\lambda,z)= \bigg[\frac{2}{32}\phi(\lambda_{0\,1},z)+ 
\frac{5}{32}\phi(\lambda_{1\,1},z)\nonumber \\ && +\frac{5}{32}\phi(\lambda_{2\,1},z) + 
\frac{1}{32}\phi(\lambda_{1\,2},z)\nonumber \\ && + \frac{5}{32}\phi(\lambda_{2\,2},z)+
\frac{14}{32}\phi(\lambda_{3\,2},z)\bigg]. 
\label{prof-combined}
\end{eqnarray}
The $17.8\%$ of $\phi_o(\lambda,z)$ in Equation~(\ref{phi-g}) contains 
contributions from both the $^{135}$Ba (6.6\%) and $^{137}$Ba (11.2\%) odd isotopes.

The reduced total source vector $\bm{\mathcal S}(\lambda, z)$
appearing in Equation~(\ref{transfer}) is defined as
\begin{eqnarray}
&&\bm{\mathcal S}(\lambda, z)=\frac{k_l(z)\bm{\mathcal S}_{l}(\lambda, z)
+\sigma_c(\lambda, z) \bm{\mathcal S}_{c}(\lambda, z)}
{k_{\rm tot}(\lambda, z)} \nonumber \\ &&
+\frac{ k_{\rm th}(\lambda, z) \bm{\mathcal S}_{\rm th}(\lambda, z)+ 
\epsilon~k_l(z)\phi_g(\lambda, z)\bm{\mathcal S}_{\rm th}(\lambda, z) }
{k_{\rm tot}(\lambda, z)},\nonumber \\
\label{s-total}
\end{eqnarray}
for a two-level atom with an unpolarized lower level. 
For the case of Ba~{\sc ii} D$_2$ it was shown by \cite{2008A&A...481..845D}
that any ground level polarization would be destroyed by elastic collisions with 
hydrogen atoms \citep[see also][]{2009A&A...493..201F}.

In Equation~(\ref{s-total}), $\bm{\mathcal S}_{\rm th}=(B_{\lambda},0)^T$, 
where $B_{\lambda}$ is
the Planck function. ${\bm{\mathcal  S}}_{l}(\lambda, z)$ is the 
line source vector
defined as
\begin{eqnarray}
{\bm{\mathcal  S}}_{l}(\lambda, z)&=&
\int_{0}^{+\infty}\frac{1}{2}
\int_{-1}^{+1} 
{\widetilde{{\bm {\mathcal R}}}(\lambda ,\lambda^{\prime}, z) }
\hat\Psi(\mu')\nonumber \\ && \times {\bm{\mathcal I}}
(\lambda', \mu', z)\,d\mu'\,d\lambda',\ \ \ \ \ 
\label{irreducible-sl}
\end{eqnarray}
with
\begin{equation}
 {\widetilde{{\bm {\mathcal R}}}(\lambda ,\lambda^{\prime}, z)}=
 0.822~\widetilde{{\bm {\mathcal R}}_{e}}(\lambda ,\lambda^{\prime}, z)
+ 0.178~\widetilde{{\bm {\mathcal R}}_{o}}(\lambda ,\lambda^{\prime}, z).
\label{comb-rm}
\end{equation}
Here $\widetilde{{\bm {\mathcal R}}_{o}}(\lambda,\lambda^{\prime}, z)$ 
is a $(2\times 2)$ diagonal matrix, which includes the effects of HFS
for the odd isotopes.
Its elements are
 $\widetilde{{\bm {\mathcal R}}_o}=$ diag $({\mathcal R}_o^{0},{\mathcal R}_o^{2})$, 
where ${\mathcal R}_o^{K}$ are the redistribution 
function components for the multipolar index $K$, containing 
both type-II and type-III redistribution of \cite{1962MNRAS.125...21H}. 
The expression for ${\mathcal R}^{K}$ is obtained by 
the quantum number replacement  $L\rightarrow J;\ \ \ J\rightarrow F;\ \ \ S\rightarrow I_s$
in Equation~(7) of \citet[][see also \citealt{2013JQSRT.115...46S} and P1]{2012ApJ...758..112S}.
In our present computations, we replace the type-III redistribution functions by
CRD functions. We have verified that both of these give nearly identical results
\citep[see also][]{1978stat.book.....M, 2012A&A...541A..24S}
and such a replacement drastically reduces the computation time.
The  redistribution matrix for the $17.8\%$ of the odd isotopes
includes the contributions from the  
individual redistribution matrices for the $^{135}$Ba and $^{137}$Ba isotopes.

$\widetilde{{\bm {\mathcal R}}_{e}}(\lambda,\lambda^{\prime}, z)$ 
is also a $(2\times 2)$ diagonal matrix for the even isotopes without HFS.
Its elements ${\mathcal R}_e^{K}$ are the redistribution functions
corresponding to the $J_a=1/2 \to J_b=3/2 \to J_f=1/2$ scattering transition.
They are obtained by setting the nuclear spin
$I_s=0$ in  $\widetilde{{\bm {\mathcal R}}_o}(\lambda,\lambda',z)$.
An expression for $\widetilde{{\bm {\mathcal R}}_{e}}(\lambda,\lambda^{\prime}, z)$
can be found in \cite{1988ApJ...334..527D} and in 
\citet[][see also \citealt{1994ApJ...432..274N},
~\citealt{2011ApJ...731..114S}]{1997A&A...328..706B} 
in the Stokes vector basis. 
It is the angle averaged versions of these quantities that are used in
our present computations. As has been demonstrated in \cite{2013MNRAS.429..275S},
the use of the angle-averaged redistribution matrix is sufficiently
accurate for all practical purposes. 
 
Like in \cite{2012ApJ...758..112S} we use the two branching ratios
defined by
\begin{eqnarray}
&& A=\frac{\Gamma_{R}}{\Gamma_{R}+\Gamma_{I}+\Gamma_{E}},\nonumber \\ && 
\!\!\!\!\!\!\!\!\!\!\!\!\!\!\!
B^{(K)}=\frac{\Gamma_{R}}{\Gamma_{R}+\Gamma_{I}+D^{(K)}}\frac{\Gamma_{E}-D^{(
K)}}{\Gamma_{R}+\Gamma_{I}+\Gamma_{E}}.
\label{a-bk}
\end{eqnarray}
$\Gamma_R$ and $\Gamma_I$ are
the radiative and inelastic collisional rates, respectively.
$\Gamma_E$ is the elastic collision rate computed from 
 \cite{1998MNRAS.300..863B}.
$D^{(K)}$ are the depolarizing 
elastic collision rates with $D^{(0)}=0$. The $D^{(2)}$ is computed
using \citep[see][]{2008A&A...481..845D,2009A&A...493..201F}
\begin{eqnarray}
&& D^{(2)}=6.82\times 10^{-9}n_{\rm H}(T/5000)^{0.40} 
\nonumber \\ && \!\!\!\!\!\!\!\!\!\!\!\!\!\!\!
+ 7.44\times10^{-9}(1/2)^{1.5}n_{\rm H}(T/5000)^{0.38}\exp(\Delta E/kT),\nonumber \\
\end{eqnarray}
where $n_{\rm H}$ is the neutral hydrogen number density, $T$ the temperature,
and $\Delta E$ the energy difference between the $^2P_{1/2}$ and $^2P_{3/2}$
fine structure levels. 
In the present treatment we neglect the collisional coupling between the 
$^2P_{3/2}$ level and the metastable $^2D_{5/2}$ level. The importance of
such collisions for the line center polarization of Ba~{\sc ii} 
D$_2$ has been pointed out by \cite{2008A&A...481..845D}, who showed that
the neglect of such collisions would lead to an overestimate of the 
line core polarization by $\sim 25\%$. This in turn would cause the 
microturbulent magnetic field $(B_{\rm turb})$ to be overestimated by $\sim 35\%$, as shown
by \cite{2009A&A...493..201F}. However, the aim of the present paper
is not to determine the value of $B_{\rm turb}$ but to explore 
the roles of PRD, HFS, quantum interferences, and the atmospheric
temperature structure in the modeling of the 
triple peak structure of the Ba~{\sc ii} D$_2$ linear polarization 
profile.

Frequency coherent scattering is assumed 
in the continuum \citep[see][]{2012A&A...541A..24S} with its source vector given by
\begin{equation}
{\bm{\mathcal  S}}_{c}(\lambda, z)= \frac{1}{2}\int_{-1}^{+1} \hat\Psi(\mu')
{\bm{\mathcal I}}(\lambda, \mu', z)\,d\mu'.
\label{irreducible-sc}
\end{equation}
The matrix $\hat\Psi$ is the Rayleigh scattering phase matrix in the 
reduced basis \citep[see][]{2007A&A...476..665F}.
The line thermalization parameter $\epsilon$ is defined by
$\epsilon={\Gamma_I}/(\Gamma_R+\Gamma_I)$. 
The Stokes vector $(I,Q)^{\rm T}$ can be computed from the 
irreducible Stokes vector $\bm{\mathcal I}$ by simple transformations
given by \citep[see][]{2007A&A...476..665F}
\begin{eqnarray}
I(\lambda,{\theta},z)&=&\mathcal{I}^{0}_{0}(\lambda,\mu,z)\nonumber \\
&& + \frac{1}{2\sqrt{2}}(3\cos^2\theta-1)
\mathcal{I}^{2}_{0}(\lambda,\mu,z),\nonumber \\
Q(\lambda,\bm{\theta},z)&=& \frac{3}{2\sqrt{2}}(1-\cos^2\theta)
\mathcal{I}^{2}_{0}(\lambda,\mu,z),\ \ \ \ 
\end{eqnarray}
where $\theta$ is the colatitude of 
the scattered ray. The scattering geometry is 
shown in Figure~1 of \cite{2011ApJ...737...95A}.
\section{Observational details}
\label{obs-details}
\begin{figure}
\centering
\includegraphics[width=7.0cm]{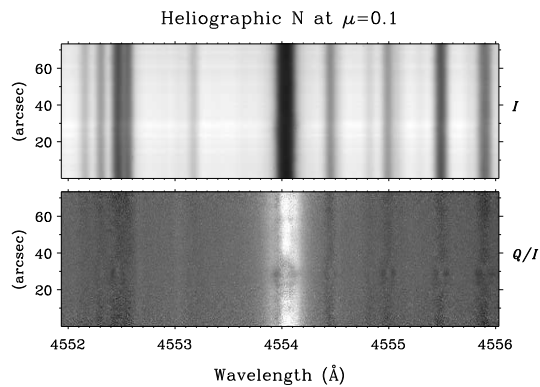}
\caption{CCD image showing the $I$ and $Q/I$ spectra of the Ba~{\sc ii} D$_2$
line at $\mu=0.1$. The observations were obtained on June 3, 2008,
with ZIMPOL-2 at the French THEMIS telescope on Tenerife.}
\label{ccd}
\end{figure}

The observed polarization profiles of the Ba~{\sc ii} D$_2$ line that
are used in the present paper for modeling purposes
were acquired by the ETH team of Stenflo on 
June 3, 2008,  using their ZIMPOL-2 imaging polarimetry system
\citep{2004A&A...422..703G} at the THEMIS telescope on Tenerife.  
Figure~{\ref{ccd}} shows the CCD 
image of the data recorded at the heliographic north pole with the
spectrograph slit placed
parallel to the limb at $\mu=0.1$. The polarization modulation was done
using Ferroelectric Liquid Crystal (FLC) modulators. 
The spectrograph slit was 1\arcsec\  wide and 70\arcsec\ long on the solar disk.
The resulting CCD image has 140 pixels in the spatial direction 
and 770 pixels in the spectral direction. 
The effective pixel size was 0.5\arcsec spatially and 5.93~m\AA\  spectrally. 
The observed profiles used to compare with the theoretical ones
have been obtained by averaging the $I$ and $Q/I$ images in Figure~{\ref{ccd}}
over the spatial interval 40\arcsec - 52\arcsec.

The recording presented in Figure~\ref{ccd} does not show much spatial 
variation along the slit, since it represents a very quiet region. However, 
recordings near magnetic regions made during the same observing campaign with ZIMPOL on THEMIS exhibit large  
spatial variations. It has long been known that all strong chromospheric 
scattering lines (like the Ca~{\sc i} 4227~\AA, Na~{\sc i} D$_2$, Sr~{\sc ii}
4079~\AA~line, etc) have such spatial variations. Our observations confirm that the 
Ba~{\sc ii} D$_2$  line is no exception, which means that it is sensitive to the 
Hanle effect like the other chromospheric lines. Observations of spatial 
variations of this line have also been carried out 
by \cite{2009A&A...501..729L} and \cite{2009ASPC..405...41R}.

\section{Modeling procedure}
\label{modeling-pro}
To model the polarization profiles of the Ba~{\sc ii} 
D$_2$ line we use a procedure similar to the
one described in \citet[][see also \citealt{2011ApJ...737...95A},
~\citealt{2010ApJ...718..988A},
~\citealt{2012A&A...541A..24S}]{2005A&A...434..713H}. 
It involves the computation of the intensity, opacity and collisional rates
from the PRD-capable MALI (Multi-level Approximate Lambda
Iteration) code developed by \citet[][referred to as the RH-code]{2001ApJ...557..389U}. 
The code solves the statistical equilibrium equation and the
unpolarized radiative transfer equation self-consistently.  
The  opacities and the collision rates thus obtained are kept fixed, 
while the reduced Stokes vector $\bm{\mathcal{I}}$ 
is computed perturbatively by solving the polarized radiative transfer equation with
the angle-averaged redistribution matrices defined in Section~{\ref{rt}}.

Such a procedure requires a model atom and a model atmosphere as inputs
to the RH-code. The details of the model atom and the atmosphere are discussed 
in the next subsections. 

\subsection{Model atom}
\label{model-atom}
Three different atom models are considered, two for the odd 
and one for the even isotope. 
The atom model for the even isotope ($^{138}$Ba) is given by the 
five levels of Figure~{\ref{level-diag}(a)},
while for the odd isotopes ($^{135}$Ba and $^{137}$Ba) the model 
is extended to include the hyperfine splitting as
described by Figure~{\ref{level-diag}(b)}.
We neglect the contribution from other less abundant even isotopes.
The wavelengths of the six hyperfine transitions
for the odd isotopes are taken from Kurucz' database
and are listed in Table~1.
These transitions are weighted with their line strengths 
given in Equation~(\ref{prof-combined})
(see Table~1). 

\subsection{Model atmosphere}
\label{model-atmos}
\begin{figure*}
\centering
\includegraphics[width=14.0cm]{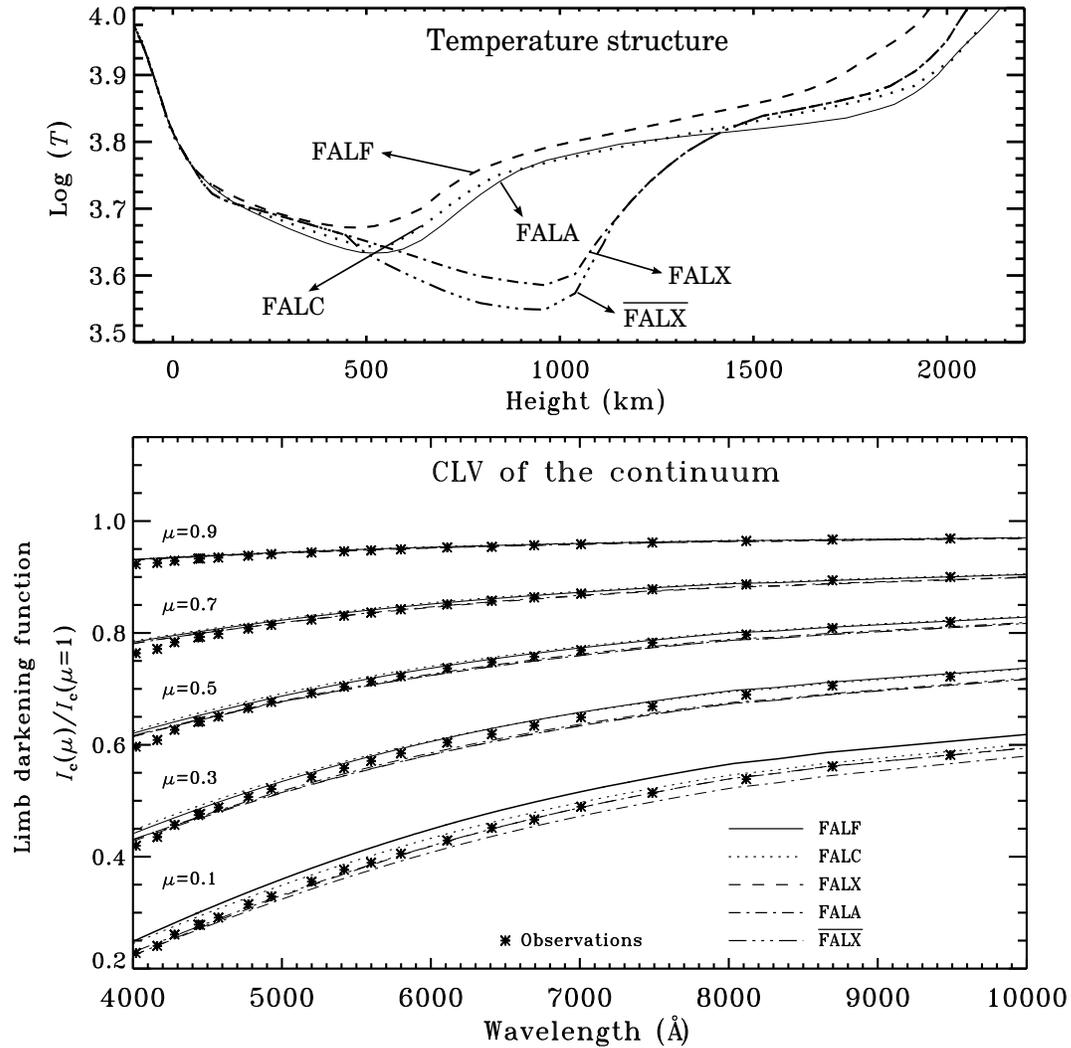}
\caption{{\it Top panel:} The temperature structure of some of the standard
model atmospheres. $\overline{\rm {FALX}}$ represents the 
model for which the temperature is reduced 
by about 300 K over a 700 km range around the height of formation
of the Ba\,{\sc ii} D$_2$ line. {\it Bottom panel:} Comparison between the
observed center-to-limb variation (CLV) of the 
continuum intensity and the predictions from different model
atmospheres including $\overline{\rm FALX}$ 
for a wide range of wavelengths from the violet to the IR 
region of the spectrum. For all the $\mu$ values the dashed and the dash-triple 
dotted lines are practically indistinguishable as the models FALX and
$\overline{\rm FALX}$ produce nearly identical fits.}
\label{clv}
\end{figure*}

We present the results computed for some of the 
standard realistic 1-D model atmospheres, like FALA, FALF, FALC \citep{1993ApJ...406..319F} 
and FALX \citep{1995itsa.conf..303A}.  
Among these four models FALF is the hottest and FALX the coolest. 
Their temperature structures are shown in the top panel of Figure~{\ref{clv}}.
However, as will be discussed below, we find that a model atmosphere
that is cooler than FALX is needed to fit the observed profiles. 
The new model, denoted $\overline{\rm FALX}$, is obtained 
by reducing the temperature of the FALX model by about 300 K in the
height range 500 -- 1200 km above the photosphere. 

We have verified that 
such a modification 
of the FALX model does not significantly affect the intensity spectra.
In contrast, the $Q/I$ spectra turn out to be very sensitive to such temperature changes.
Like in \cite{2012A&A...541A..24S}, we test the  $\overline{\rm FALX}$
atmosphere by computing the limb darkening function for a range of wavelengths 
and $\mu$ values and compare it with the observed data from 
\cite{1994SoPh..153...91N}. This is shown in the bottom panel of Figure~{\ref{clv}}.
One can see that $\overline{\rm FALX}$ and the standard FALX fit the observed
center-to-limb variation equally well. Therefore small modifications
of  the temperature structure to achieve a good fit to the observed
$Q/I$ profile can be made without affecting the model constraints
imposed by the intensity spectrum.

\section{Results}
\label{results}
In the following we discuss the modeling details and
the need for a model atmosphere that is cooler than FALX.
This helps us to evaluate the temperature sensitivity
of the Ba~{\sc ii} D$_2$ line and its usefulness for magnetic-field
diagnostics. In addition we demonstrate the profound role that PRD
plays for the formation of the polarized line profile. 

\subsection{Modeling the Ba~{\sc ii} D$_2$ line profile}
\label{mod-obs}
\subsubsection{Modeling details}
\label{mod-details}
From the three Ba~{\sc ii} atom models described in Section~{\ref{model-atom}}
we obtain three sets of physical quantities
(two for the odd isotopes and one for the even isotope) from the RH code. 
These quantities include line opacity, line emissivity, 
continuum absorption coefficient, continuum emissivity, continuum
scattering coefficient, and the mean intensity. 
{The mathematical expressions used to compute these various quantities
for the even isotopes are given in \cite{2001ApJ...557..389U}. 
For the odd isotopes, the profile functions in these expressions
are replaced by $\phi_o(\lambda,z)$ defined by Equation~(\ref{prof-combined}).}

The three sets of quantities  
are then combined in the ratio of their respective isotope abundances
and subsequently used as inputs to the polarization code.

\subsubsection{Temperature sensitivity}
\begin{figure*}
\centering
\includegraphics[width=10.0cm]{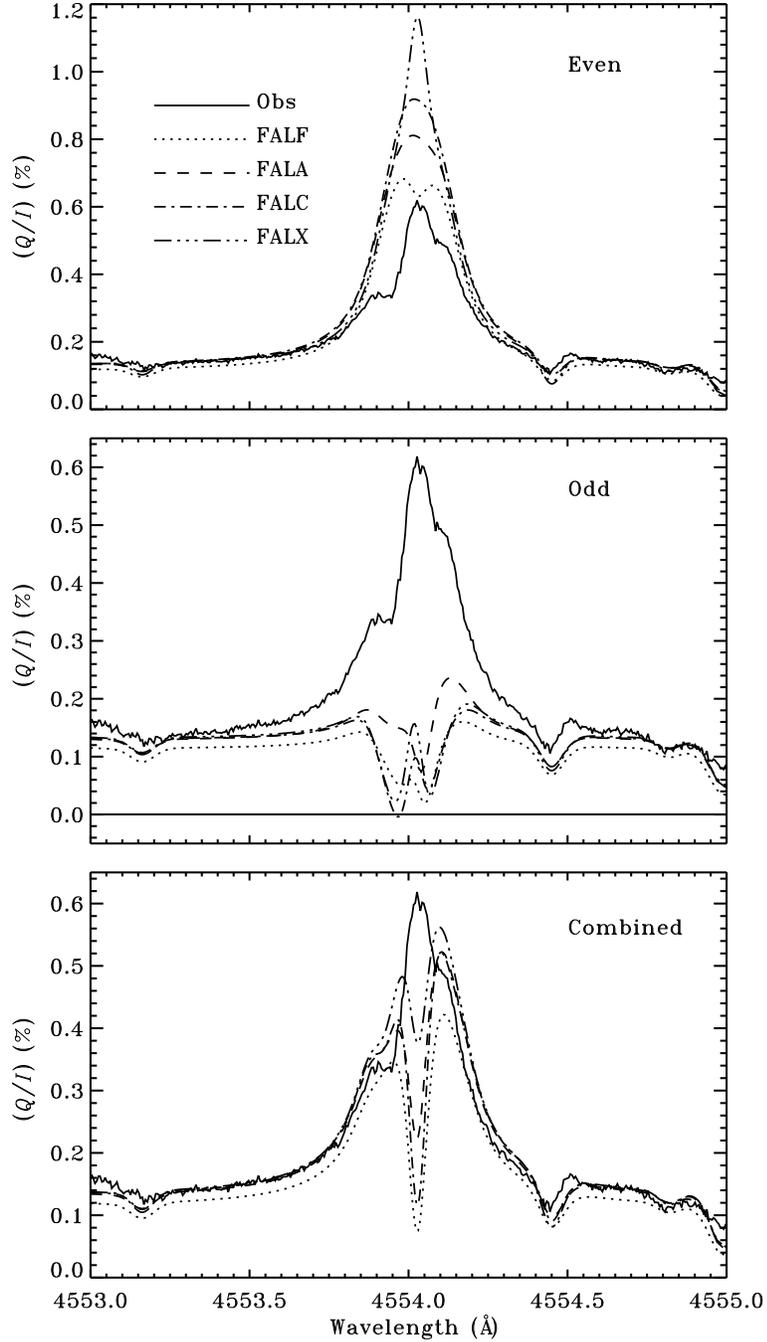}
\caption{Comparison between the observed $Q/I$ profile and the theoretical profiles
for some of the standard model atmospheres, separately displayed for
the even, odd, and combined even-odd cases. The theoretical profiles
represent the non-magnetic case and have been smeared with a Gaussian
having a full width at half maximum (FWHM) of 70~m\AA\ to account for
instrumental and macro-turbulent broadening.}
\label{all-models}
\end{figure*}

The polarization profiles thus computed 
for the various model atmospheres are shown in
Figure~\ref{all-models}, displayed separately  for the even, odd, 
and combined even-odd cases in three different panels. 
The Stokes $Q/I$ profiles in Figure~\ref{all-models} are 
computed by setting the total abundance of Ba in the Sun equal
to the abundance of even isotopes in the first panel; the abundance of odd isotopes
in the second panel; and a fractional abundance of even (82\%) and 
odd (18\%) isotopes in the third panel. The profiles in the first panel can be compared 
to the results presented in  Figure~6 of \cite{2009A&A...493..201F}.
As seen from the first panel, the amplitude of the central peak
for the even isotopes is very sensitive to the temperature structure 
of the model atmosphere
in contradiction with the conclusions of 
\cite{2009A&A...493..201F}. Also in their paper, the amplitude of the central peak 
obtained from the FALC model is larger than the one obtained from
FALX, which is opposite to our findings (although it could be that the version
of the FALC model they used is not identical to the one that we have
used). However, the profile computed with the FALX model in first panel of 
Figure~\ref{all-models} for the even isotopes is in good agreement
with the one given in their paper.

The profiles in the second panel of Figure~\ref{all-models}, which
represent the odd isotopes, also exhibit a similar large sensitivity
to the choice of model atmosphere.  
Therefore the combined even-odd isotopes profiles in the third panel 
are also very sensitive to the temperature structure. 

For the sake of clarity, let us point out that the combined $Q/I$
profiles in the bottom panel of Figure~{\ref{all-models}} differ profoundly from 
what one would obtain from a linear superposition of the 
corresponding profiles for the even and odd isotopes individually in the 
two other panels, in proportion to their isotope ratios. The
reason is that the combination is highly non-linear, since
the lines are formed in an optically thick medium
(namely the radiative transfer effects). While 
the opacities and redistribution matrices are combined in
a linear way as described by Equations~(\ref{phi-g}) and (\ref{comb-rm}), the 
even and odd isotopes blend with each other in the radiative 
transfer process, which makes the combination as it appears
in the emergent spectrum highly non-linear.

The drastic depolarization of $Q/I$ in the line core has its origin in the 
polarizability factor $W_2$ of the odd isotopes. It is well known that the 
trough like suppression of $W_2$ in the line core for Ba~{\sc ii} is formed
due to hyperfine structure for the odd isotopes \citep[see][]{1997A&A...324..344S}.
In our radiative transfer calculations we have a superposition
(in the proportion of the isotope ratios), of the trough-like scattering 
opacity of the odd isotopes, with the peak-like scattering opacity of 
the even isotopes. The shape of the $Q/I$ profile depends on the
details of radiative transfer and PRD (see Equations~(\ref{phi-g})-(\ref{comb-rm}))
namely on how these two scattering opacities non-linearly blend to produce the net 
result for the emergent radiation in the optically thick cases.

\subsubsection{Need for the $\overline{\rm FALX}$ model}
As seen from the last panel of Figure~\ref{all-models}, 
the central peak is not well reproduced
by any of the standard model atmospheres. All the models produce a dip
at line center.  Such a central dip is commonly due to the effects of
PRD, caused by the properties of the type-II frequency
redistribution. In the case of Ba~{\sc ii} D$_2$, 
the contribution to this central dip comes mainly from the even isotopes, 
as shown in Figure~\ref{r-ii-dip}. The three rows in this figure 
represent the even, odd and combined even-odd cases, respectively, 
for the FALX model atmosphere. The first column 
shows the profiles computed with only type-II redistribution, the 
second column those computed with CRD only.
The CRD profiles are obtained by setting the branching
ratios $A=0$ and $B^{(K)}=(1 - \epsilon)$. None of the CRD profiles
shows a central dip. 

The occurrence and nature of this central dip has been explored
in detail in \cite{2005A&A...434..713H}, who showed that its magnitude is strongly 
dependent on the choice of atmospheric parameters. 
This behavior is also evident from Figure~\ref{all-models}. 
The cooler the atmosphere, the smaller is the central dip. 
The dip is often smoothed out by instrumental and macro-turbulent
broadening. However, the profiles in Figure~\ref{all-models} have
already been smeared with a Gaussian 
function having a full width at half maximum (FWHM) of 70~m\AA. The
dip could be suppressed by additional smearing, but such large
smearing would also suppress the observed side peaks of the odd
isotopes and would make the intensity profile inconsistent with the
observed one. The value 70~m\AA\ has been chosen to optimize the fit,
but it is also consistent with what we expect based on the observing
parameters and turbulence in the chromosphere. 

The failure of all the tried standard model atmospheres therefore
leads us to introduce a new model with a modified temperature
structure, which is cooler
than the standard FALX model. The details of the new cooler 
 $\overline{\rm FALX}$ model has been given in Section~{\ref{model-atmos}}.
This new model atmosphere succeeds in giving a good fit
to  both the intensity and the polarization profiles, as shown in Figure~\ref{final-fit}.
To simulate the effects of spectrograph stray light on the intensity
and polarization profiles we have applied 
a spectrally flat unpolarized background of 4\%\ of the continuum intensity level
to the theoretical $(I,Q/I)$ profiles. For a good $Q/I$ line center fit, we find
that it is necessary to include Hanle
depolarization from a non-zero magnetic field. 
Our theoretical profiles are based on a  micro-turbulent magnetic
field of strength $B_{\rm turb}$ with an isotropic angular 
distribution.  Our best fit to the $Q/I$ profile 
corresponds to a field strength of $B_{\rm turb}$ = 2~G.

\subsection{The importance of PRD}
\label{crd}
\begin{figure*}
\centering
\includegraphics[width=13.0cm]{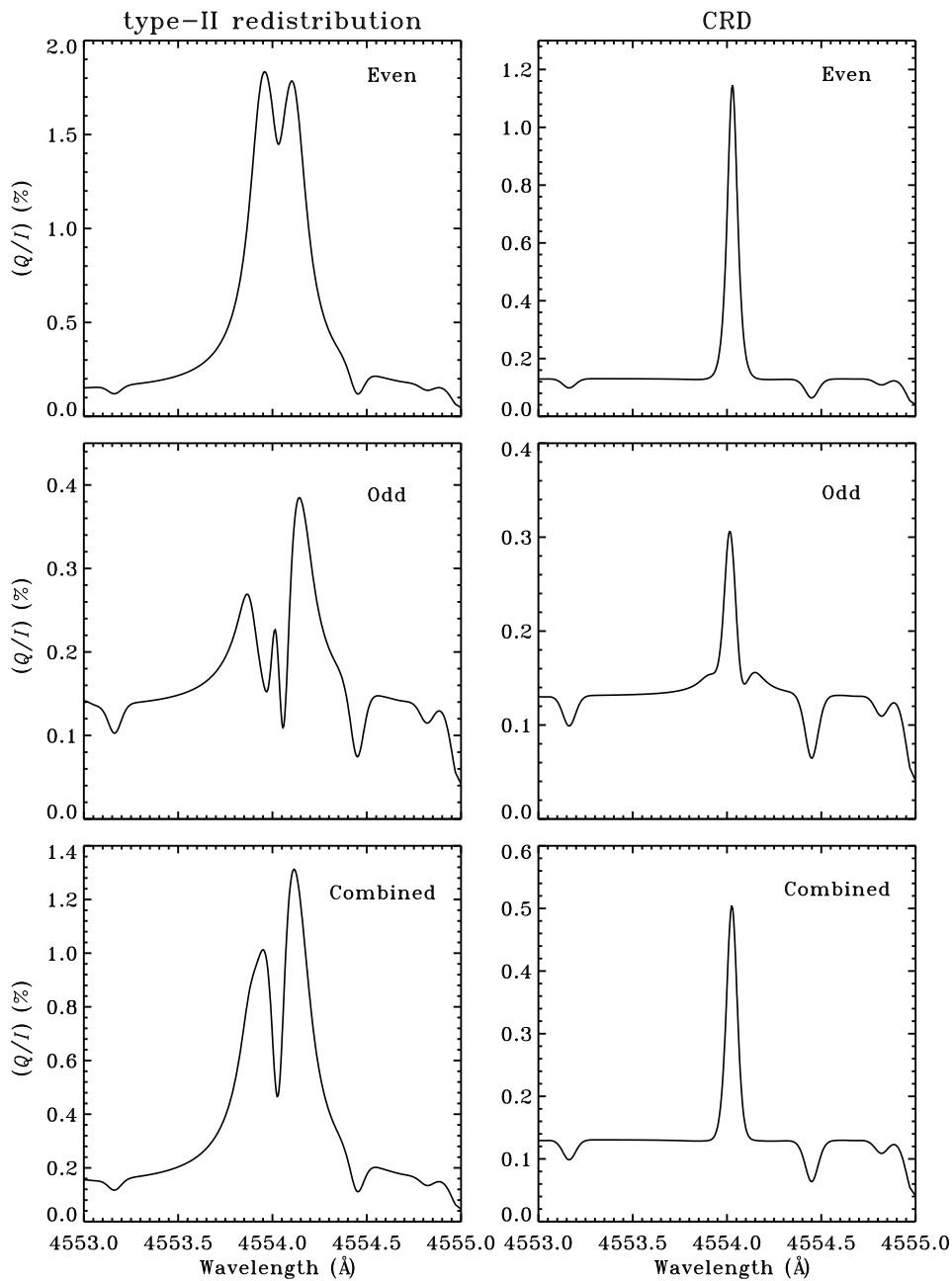}
\caption{Theoretical $Q/I$ profiles computed for the
  non-magnetic FALX model for the even (first row), odd 
(second row), and combined even-odd (third row) isotopes, with
only type-II frequency redistribution (first column) and only complete
frequency redistribution CRD (second column). A prominent central dip
is present for the type-II redistribution profiles although they have
been smeared with a Gaussian having FWHM = 70~m\AA.}
\label{r-ii-dip}
\end{figure*}

\begin{figure*}
\centering
\includegraphics[width=9.0cm]{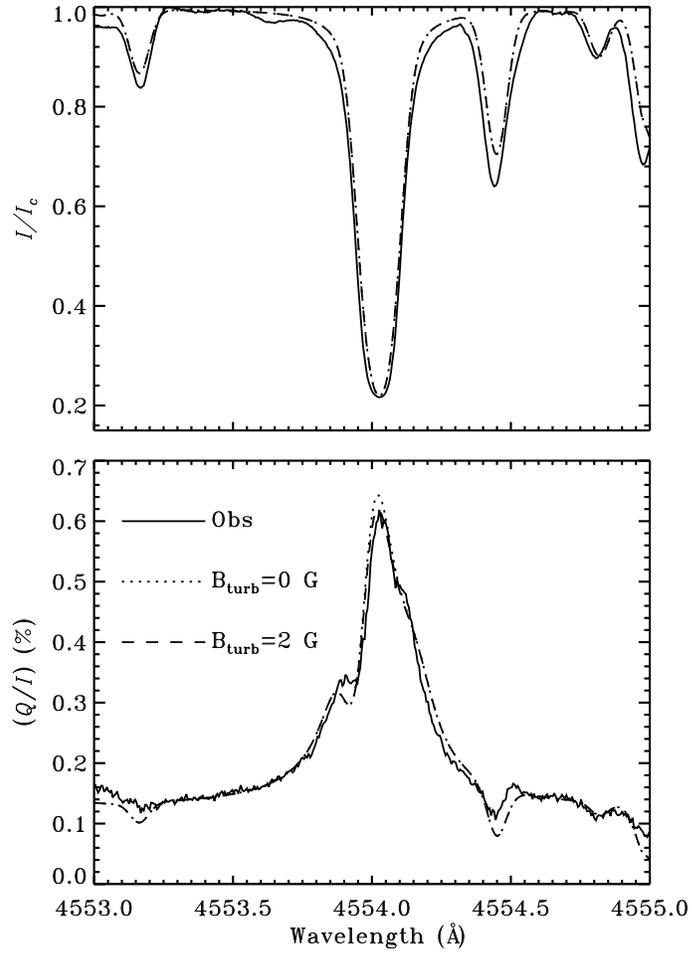}
\caption{Fit to the observed profile using the $\overline{\rm {FALX}}$
model with and without a micro-turbulent magnetic field $B_{\rm turb}$.
The theoretical profiles have been smeared using a Gaussian with FWHM = 70~m\AA.}
\label{final-fit}
\end{figure*}

\begin{figure*}
\centering
\includegraphics[width=9.0cm]{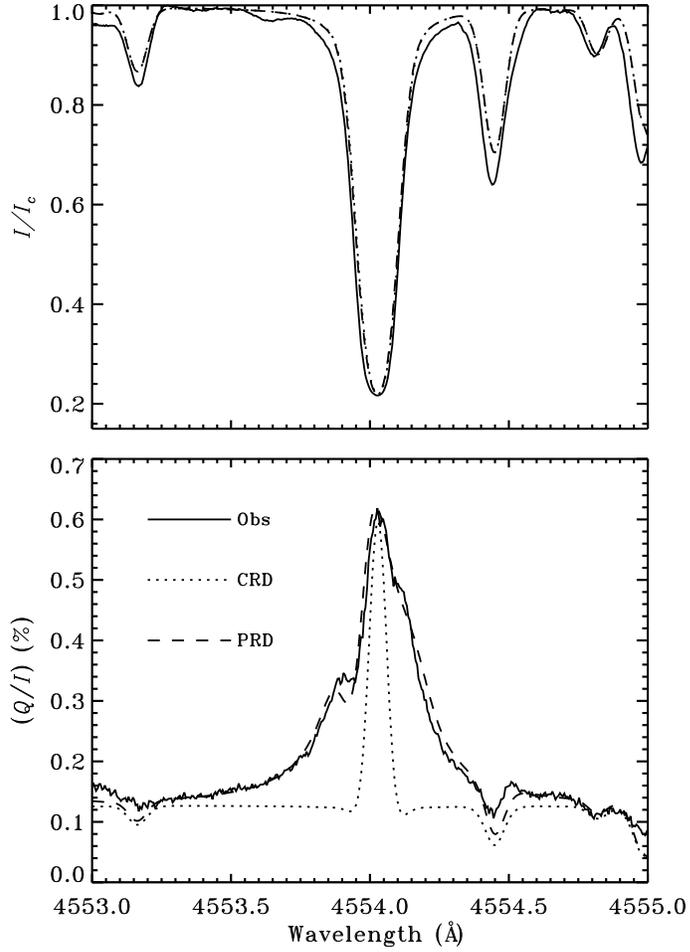}
\caption{Comparison between the observed Stokes profiles and the profiles
  computed with CRD (dotted line) and 
PRD (dashed line) for the  $\overline{\rm FALX}$ model.
The theoretical profiles have been smeared with a Gaussian having FWHM = 70~m\AA.
The strength of the micro-turbulent magnetic field 
$B_{\rm turb}$ has been chosen to be 2~G for the PRD and 5~G for the CRD profiles.
The dashed line in this figure is same as the dashed line in Figure~{\ref{final-fit}}.}
\label{crd-prd-falxd}
\end{figure*}

The importance of PRD in modeling the Ba\,{\sc ii} D$_2$ line has 
already been demonstrated in \cite{2009A&A...493..201F}, although by only considering 
the even isotopes. Figure~{\ref{r-ii-dip}} 
demonstrates the importance of PRD for both
the odd and the even isotopes. 
As seen from the second column of this figure, the $Q/I$ profiles for the 
odd isotopes when computed exclusively in CRD 
do not produce any side peaks, while the profiles computed 
with type-II redistribution
exhibits such peaks. Also, by comparing the $Q/I$ profiles 
for the even isotopes in the first row, we see that CRD 
fails to generate the needed line wing polarization. 
A comparison between the observed profiles and the theoretical profiles 
based on CRD alone (dotted line) and on full PRD (dashed line) 
for the $\overline{\rm {FALX}}$ model is shown in Figure~{\ref{crd-prd-falxd}}.
While the intensity profile can be fitted well using either PRD or
CRD, the polarization profile cannot be fitted at all with CRD alone.
PRD is therefore essential to model the $Q/I$ profiles of the Ba~{\sc ii} D$_2$ line.  

\section{Conclusions}
\label{conclusions}
In the present paper we have for the first time tried to model 
the polarization profiles of the Ba~{\sc ii} D$_2$ line by taking full 
account of PRD, radiative transfer, and HFS effects. 
We use the theory of $F$-state interference developed in P1
in combination with different atom models representing different
isotopes of Ba~{\sc ii} and various choices of model atmospheres.
Applications of the well known
standard model atmospheres FALF, FALC, FALA, and FALX fail to reproduce
the central peak, and instead produce a central dip mainly due to 
PRD effects. We have shown that in the case of Ba~{\sc ii} D$_2$ the
central dip is reduced by 
lowering the temperature of the atmospheric model. We can therefore achieve
a good fit to the observed polarization profile by slightly reducing
the temperature of the FALX model.

In modeling the Ba~{\sc ii} D$_2$ line we account for the depolarizing effects
of elastic collisions with hydrogen atoms but neglect the alignment
transfer between the $^2P_{3/2}$ level and the metastable $^2D_{5/2}$ level.
It has been shown by 
\cite{2008A&A...481..845D} that this alignment transfer affects the 
line center polarization and is needed for magnetic field
diagnostics. The purpose of the present paper is however not to
determine magnetic fields but to clarify the physics of line
formation. 

We demonstrate that PRD is essential to reproduce the 
triple peak structure and the line wing polarization of the Ba~{\sc ii}
D$_2$ line, but find that the line center polarization is very
sensitive to the  temperature structure of the atmosphere, which 
contradicts the conclusions of \cite{2009A&A...493..201F}, 
who find that the barium line is temperature insensitive and therefore suitable for 
Hanle diagnostics. This contradiction illustrates that a full PRD 
treatment as done in the present paper, including the contributions 
from both the even and odd isotopes, is necessary to bring out the
correct temperature dependence of the line.
The large temperature sensitivity of the Ba~{\sc ii} D$_2$ 
line makes it rather unsuited for magnetic-field
diagnostics, since there is no known straightforward way to separate
the temperature and magnetic-field effects for this line.

\acknowledgments
We would like to thank Dr.~Michele Bianda for useful discussions.


\begin{thebibliography}{32}
\expandafter\ifx\csname natexlab\endcsname\relax\def\natexlab#1{#1}\fi

\bibitem[{{Anusha} {et~al.}(2010){Anusha}, {Nagendra}, {Stenflo}, {Bianda},
  {Sampoorna}, {Frisch}, {Holzreuter}, \& {Ramelli}}]{2010ApJ...718..988A}
{Anusha}, L.~S., {Nagendra}, K.~N., {Stenflo}, J.~O., {Bianda}, M.,
  {Sampoorna}, M., {Frisch}, H., {Holzreuter}, R., \& {Ramelli}, R. 2010, \apj,
  718, 988

\bibitem[{{Anusha} {et~al.}(2011){Anusha}, {Nagendra}, {Bianda}, {Stenflo},
  {Holzreuter}, {Sampoorna}, {Frisch}, {Ramelli}, \&
  {Smitha}}]{2011ApJ...737...95A}
{Anusha}, L.~S., {et~al.} 2011, \apj, 737, 95

\bibitem[{{Asplund} {et~al.}(2009){Asplund}, {Grevesse}, {Sauval}, \&
  {Scott}}]{2009ARA&A..47..481A}
{Asplund}, M., {Grevesse}, N., {Sauval}, A.~J., \& {Scott}, P. 2009, \araa, 47,
  481

\bibitem[{{Avrett}(1995)}]{1995itsa.conf..303A}
{Avrett}, E.~H. 1995, in Infrared tools for solar astrophysics: What's next?,
  ed. J.~R. {Kuhn} \& M.~J. {Penn}, 303--311

\bibitem[{{Barklem} \& {O'Mara}(1998)}]{1998MNRAS.300..863B}
{Barklem}, P.~S., \& {O'Mara}, B.~J. 1998, \mnras, 300, 863

\bibitem[{{Belluzzi} {et~al.}(2007){Belluzzi}, {Trujillo Bueno}, \& {Landi
  Degl'Innocenti}}]{2007ApJ...666..588B}
{Belluzzi}, L., {Trujillo Bueno}, J., \& {Landi Degl'Innocenti}, E. 2007, \apj,
  666, 588

\bibitem[{{Bommier}(1997)}]{1997A&A...328..706B}
{Bommier}, V. 1997, \aap, 328, 706

\bibitem[{{Derouich}(2008)}]{2008A&A...481..845D}
{Derouich}, M. 2008, \aap, 481, 845

\bibitem[{{Domke} \& {Hubeny}(1988)}]{1988ApJ...334..527D}
{Domke}, H., \& {Hubeny}, I. 1988, \apj, 334, 527

\bibitem[{{Faurobert} {et~al.}(2009){Faurobert}, {Derouich}, {Bommier}, \&
  {Arnaud}}]{2009A&A...493..201F}
{Faurobert}, M., {Derouich}, M., {Bommier}, V., \& {Arnaud}, J. 2009, \aap,
  493, 201

\bibitem[{{Fontenla} {et~al.}(1993){Fontenla}, {Avrett}, \&
  {Loeser}}]{1993ApJ...406..319F}
{Fontenla}, J.~M., {Avrett}, E.~H., \& {Loeser}, R. 1993, \apj, 406, 319

\bibitem[{{Frisch}(2007)}]{2007A&A...476..665F}
{Frisch}, H. 2007, \aap, 476, 665

\bibitem[{{Gandorfer} {et~al.}(2004){Gandorfer}, {Steiner}, {Aebersold},
  {Egger}, {Feller}, {Gisler}, {Hagenbuch}, \& {Stenflo}}]{2004A&A...422..703G}
{Gandorfer}, A.~M., {Steiner}, H.~P.~P.~P., {Aebersold}, F., {Egger}, U.,
  {Feller}, A., {Gisler}, D., {Hagenbuch}, S., \& {Stenflo}, J.~O. 2004, \aap,
  422, 703

\bibitem[{{Holweger} \& {Mueller}(1974)}]{1974SoPh...39...19H}
{Holweger}, H., \& {Mueller}, E.~A. 1974, \solphys, 39, 19

\bibitem[{{Holzreuter} {et~al.}(2005){Holzreuter}, {Fluri}, \&
  {Stenflo}}]{2005A&A...434..713H}
{Holzreuter}, R., {Fluri}, D.~M., \& {Stenflo}, J.~O. 2005, \aap, 434, 713

\bibitem[{{Hummer}(1962)}]{1962MNRAS.125...21H}
{Hummer}, D.~G. 1962, \mnras, 125, 21

\bibitem[{{L{\'o}pez Ariste} {et~al.}(2009){L{\'o}pez Ariste}, {Asensio Ramos},
  {Manso Sainz}, {Derouich}, \& {Gelly}}]{2009A&A...501..729L}
{L{\'o}pez Ariste}, A., {Asensio Ramos}, A., {Manso Sainz}, R., {Derouich}, M.,
  \& {Gelly}, B. 2009, \aap, 501, 729

\bibitem[{{Mihalas}(1978)}]{1978stat.book.....M}
{Mihalas}, D. 1978, {Stellar atmospheres /2nd edition/}

\bibitem[{{Nagendra}(1994)}]{1994ApJ...432..274N}
{Nagendra}, K.~N. 1994, \apj, 432, 274

\bibitem[{{Neckel} \& {Labs}(1994)}]{1994SoPh..153...91N}
{Neckel}, H., \& {Labs}, D. 1994, \solphys, 153, 91

\bibitem[{{Ramelli} {et~al.}(2009){Ramelli}, {Bianda}, {Trujillo Bueno},
  {Belluzzi}, \& {Landi Degl'Innocenti}}]{2009ASPC..405...41R}
{Ramelli}, R., {Bianda}, M., {Trujillo Bueno}, J., {Belluzzi}, L., \& {Landi
  Degl'Innocenti}, E. 2009, in Astronomical Society of the Pacific Conference
  Series, Vol. 405, Solar Polarization 5: In Honor of Jan Stenflo, ed. S.~V.
  {Berdyugina}, K.~N. {Nagendra}, \& R.~{Ramelli}, 41

\bibitem[{{Rutten}(1978)}]{1978SoPh...56..237R}
{Rutten}, R.~J. 1978, \solphys, 56, 237

\bibitem[{{Sampoorna}(2011)}]{2011ApJ...731..114S}
{Sampoorna}, M. 2011, \apj, 731, 114

\bibitem[{{Smitha} {et~al.}(2013){Smitha}, {Nagendra}, {Sampoorna}, \&
  {Stenflo}}]{2013JQSRT.115...46S}
{Smitha}, H.~N., {Nagendra}, K.~N., {Sampoorna}, M., \& {Stenflo}, J.~O. 2013,
  \jqsrt, 115, 46

\bibitem[{{Smitha} {et~al.}(2012{\natexlab{a}}){Smitha}, {Nagendra}, {Stenflo},
  {Bianda}, {Sampoorna}, {Ramelli}, \& {Anusha}}]{2012A&A...541A..24S}
{Smitha}, H.~N., {Nagendra}, K.~N., {Stenflo}, J.~O., {Bianda}, M.,
  {Sampoorna}, M., {Ramelli}, R., \& {Anusha}, L.~S. 2012{\natexlab{a}}, \aap,
  541, A24

\bibitem[{{Smitha} {et~al.}(2011){Smitha}, {Sampoorna}, {Nagendra}, \&
  {Stenflo}}]{2011ApJ...733....4S}
{Smitha}, H.~N., {Sampoorna}, M., {Nagendra}, K.~N., \& {Stenflo}, J.~O. 2011,
  \apj, 733, 4

\bibitem[{{Smitha} {et~al.}(2012{\natexlab{b}}){Smitha}, {Sowmya}, {Nagendra},
  {Sampoorna}, \& {Stenflo}}]{2012ApJ...758..112S}
{Smitha}, H.~N., {Sowmya}, K., {Nagendra}, K.~N., {Sampoorna}, M., \&
  {Stenflo}, J.~O. 2012{\natexlab{b}}, \apj, 758, 112

\bibitem[{{Stenflo}(1980)}]{1980A&A....84...68S}
{Stenflo}, J.~O. 1980, \aap, 84, 68

\bibitem[{{Stenflo}(1997)}]{1997A&A...324..344S}
---. 1997, \aap, 324, 344

\bibitem[{{Stenflo} \& {Keller}(1997)}]{1997A&A...321..927S}
{Stenflo}, J.~O., \& {Keller}, C.~U. 1997, \aap, 321, 927

\bibitem[{{Supriya} {et~al.}(2013){Supriya}, {Smitha}, {Nagendra}, {Ravindra},
  \& {Sampoorna}}]{2013MNRAS.429..275S}
{Supriya}, H.~D., {Smitha}, H.~N., {Nagendra}, K.~N., {Ravindra}, B., \&
  {Sampoorna}, M. 2013, \mnras, 429, 275

\bibitem[{{Uitenbroek}(2001)}]{2001ApJ...557..389U}
{Uitenbroek}, H. 2001, \apj, 557, 389

\end{thebibliography}


\begin{table}
\caption{Wavelengths (\AA) of the hyperfine transitions for the odd isotopes of Ba~{\sc ii}}
\vspace{5pt}
\label{table-1}
\centering
\begin{tabular}{cccccc}
\hline
$F_a$ & $F_b$& $^{135}$Ba & $^{137}$Ba & Line strength \\
\hline 
1 & 0 &4553.999 & 4553.995 & 0.15625\\
1 & 1 &4554.001 & 4553.997 & 0.06250\\
1 & 2 &4554.001 & 4553.998 & 0.15625\\
2 & 1 &4554.046 & 4554.049 & 0.43750\\
2 & 2 &4554.059 & 4554.051 & 0.15625 \\
2 & 3 &4554.050 & 4554.052 & 0.03125\\

\hline

\hline
\end{tabular}
\end{table}

\end{document}